\documentclass{appolb}
\usepackage{graphicx}
\usepackage{scalefnt}

\begin{document}

\title{Phenomenology of Axial-Vector Mesons from an Extended
Linear Sigma Model \thanks{%
Presented by D.~Parganlija at Excited QCD 2012, Peniche/Portugal, 6.-12.\ May 2012} }
\author{Denis Parganlija$^{a,b}$, P\'{e}ter Kov\'{a}cs$^{b,c}$, Gy\"{o}rgy Wolf$^c$, Francesco Giacosa$^{b}$, and Dirk H. Rischke$^{b,d}$ 
\address{$^a$Institute for Theoretical Physics, Vienna University of Technology,
Wiedner Hauptstr.\ 8-10, A--1040 Vienna, Austria\\ $^b$Institute for Theoretical Physics, Goethe University,
Max-von-Laue-Str.\ 1, D--60438 Frankfurt am Main, Germany\\ $^c$Research Institute for
Particle and Nuclear Physics of the Hungarian Academy of Sciences, H-1525
Budapest, Hungary and \\ $^d$Frankfurt Institute for Advanced Studies, Goethe University,
Ruth-Moufang-Str.\ 1, D--60438 Frankfurt am Main, Germany} \\
}
\maketitle

\begin{abstract}
We discuss the phenomenology of the axial-vector mesons within a three-flavour Linear Sigma Model containing scalar,
pseudoscalar, vector and axial-vector degrees of freedom.
\end{abstract}



\section{Introduction}

A correct description of axial-vector mesons is important in Quantum Chromodynamics
(QCD) for several reasons. The lightest measured axial-vector meson -- the $%
a_1(1260)$ resonance -- is experimentally ambiguous already in vacuum:
according to listings of the Particle Data Group (PDG \cite{PDG}), the decay
width of $a_1(1260)$ possesses values between 250 MeV and 600 MeV. There is only one class of
mesons distinct from axial-vectors with similar values of decay widths: the
scalar mesons, the theoretical and experimental description of which is
famously ambiguous (see, e.g., Refs.\ \cite{Leutwyler, Doktorarbeit, Paper1, Paper3}). The large decay width
renders experimental as well as theoretical determination of the
properties of $a_1(1260)$ rather problematic.\newline
Additionally, it is well-known that the QCD Lagrangian with $N_f$ massless quark
flavours possesses an exact $SU(N_f)_A$ axial symmetry. Consequently,
a conserved axial-vector current of the form $\bar{q}_{f}\gamma^{\mu}%
\gamma_{5}t^{i}q_{f}$ arises, where $q_{f}$ denotes a quark flavour, $%
\gamma^{\mu}$ and $\gamma_{5}$ are Dirac matrices and $t^i$ represent
generators of the chiral $SU(N_f)_L \times SU(N_f)_R$ chiral group with $N_f$
flavours, $i= 1,..., N_f^2-1$. An axial rotation of this current leads to a conserved vector
current of the form $\bar{q}_{f} \gamma^{\mu} t^{i} q_{f}$. Identifying
(putatively) the latter two currents with the $a_1(1260)$ and $\rho(770)$
mesons, respectively, leads to the assertion that the latter two resonances should be degenerate in vacuum. The opposite is observed due to
the Spontaneous Breaking of the Chiral Symmetry \cite{refssb}; however, the mentioned
degeneration is expected to be restored at finite temperatures and densities
(the so-called chiral transition). A viable theoretical description
of this postulated high-temperature phenomenon requires a satisfactory
description of axial-vectors already in vacuum -- and such a description is
an objective of this article. (Note that claims have been made \cite{Glozman} that
the vector current is actually a mixture of two rotated chiral-partner
fields, an axial-vector and a pseudovector one. As a first approximation, we will neglect
this possibility.)

In this article, we present a Linear Sigma Model containing scalar,
pseudoscalar, vector and axial-vector mesons both in the non-strange and
strange sectors (extended Linear Sigma Model or eLSM\cite{Doktorarbeit, Paper3, LesHouches, Madrid, Stockholm}). 
The model contains
only $\bar q q$ states \cite{Doktorarbeit, Paper1, Paper3} rendering it appropriate to study not
only general features (masses/decays) of mesons but also their structure --
if a physical resonance can be accommodated within our model, then it possesses the $%
\bar q q$ structure. This criterion is important for scalars \cite%
{Paper1,LesHouches} but also for axial-vectors considering the claims that,
e.g., $a_1(1260)$ represents a meson-meson molecule rather than a quarkonium 
\cite{Oset}. Consequently, this paper will consider the issue whether the $a_1(1260)$
meson can be accommodated within eLSM, i.e., if $a_1$ can be described as
(predominantly) a $\bar q q$ state.

The outline of the paper is as follows. In Sec.\ \ref{Model} we present the
three-flavour Lagrangian with vector and axial-vector mesons. Consequences
of a global fit of observables for all states present in the model except
scalar isosinglets and $K_1$ are presented in Sec.\ \ref%
{Results}. We provide our conclusions in Sec.\ \ref{Conclusions}.

\section{The Model}

\label{Model}

The Lagrangian of the Extended Linear Sigma Model with the chiral $U(3)_{L}\times
U(3)_{R}$ symmetry (eLSM) reads \cite{Doktorarbeit,Paper3,LesHouches,Madrid,Stockholm}: 
\begin{eqnarray}
&&%
\mbox{\fontsize{9}{8}\selectfont $
\lefteqn{\mathcal{L}=\mathrm{Tr}[(D^{\mu }\Phi )^{\dagger }(D^{\mu
}\Phi)]-m_{0}^{2}\mathrm{Tr}(\Phi ^{\dagger }\Phi )-\lambda
_{1}[\mathrm{Tr}(\Phi^{\dagger }\Phi )]^{2}-\lambda _{2}\mathrm{Tr}(\Phi
^{\dagger }\Phi )^{2}} $}  \nonumber \\
&&%
\mbox{\fontsize{9}{8}\selectfont $ - \, \frac{1}{4}\mathrm{Tr}[(L^{\mu
\nu })^{2}+(R^{\mu \nu })^{2}]+\mathrm{Tr}\left[ \left(
\frac{m_{1}^{2}}{2}+\Delta \right) (L^{\mu })^{2}+(R^{\mu})^{2}\right]
+\mathrm{Tr}[H(\Phi +\Phi ^{\dagger })] $}  \nonumber \\
&&%
\mbox{\fontsize{9}{8}\selectfont $ + \,
c_{1}(\det\Phi-\det\Phi^{\dagger})^{2}+i\frac{g_{2}}{2}(\mathrm{Tr}\{L_{\mu
\nu }[L^{\mu },L^{\nu }]\}+\mathrm{Tr}\{R_{\mu \nu }[R^{\mu },R^{\nu }]\}) $}
\nonumber \\
&&%
\mbox{\fontsize{9}{8}\selectfont $ +\, \frac{h_{1}}{2}\mathrm{Tr}(\Phi
^{\dagger }\Phi )\mathrm{Tr}[(L^{\mu})^{2}+(R^{\mu
})^{2}]+h_{2}\mathrm{Tr}[\vert\Phi R^{\mu }\vert^{2}+\vert L^{\mu
}\Phi\vert^{2}]+2h_{3}\mathrm{Tr}(\Phi R_{\mu }\Phi ^{\dagger }L^{\mu }) $}
\label{Lagrangian}
\end{eqnarray}%
where 
\begin{equation}
\scalefont{0.81}\Phi =\frac{1}{\sqrt{2}}\left( 
\begin{array}{ccc}
\frac{(\sigma _{N}+a_{0}^{0})+i(\eta _{N}+\pi ^{0})}{\sqrt{2}} & 
a_{0}^{+}+i\pi ^{+} & K_{0}^{\star +}+iK^{+} \\ 
a_{0}^{-}+i\pi ^{-} & \frac{(\sigma _{N}-a_{0}^{0})+i(\eta _{N}-\pi ^{0})}{%
\sqrt{2}} & K_{0}^{\star 0}+iK^{0} \\ 
K_{0}^{\star -}+iK^{-} & {\bar{K}_{0}^{\star 0}}+i{\bar{K}^{0}} & \sigma
_{S}+i\eta _{S}%
\end{array}%
\right) \normalsize \label{Phi}
\end{equation}%
is a matrix containing the scalar and pseudoscalar degrees of freedom, $%
L^{\mu }=V^{\mu }+A^{\mu }$ and $R^{\mu }=V^{\mu }-A^{\mu }$ are,
respectively, the left-handed and the right-handed matrices containing
vector and axial-vector degrees of freedom with 
\begin{equation}
\scalefont{0.81}V^{\mu }=\frac{1}{\sqrt{2}}\left( 
\begin{array}{ccc}
\frac{\omega _{N}+\rho ^{0}}{\sqrt{2}} & \rho ^{+} & K^{\star +} \\ 
\rho ^{-} & \frac{\omega _{N}-\rho ^{0}}{\sqrt{2}} & K^{\star 0} \\ 
K^{\star -} & {\bar{K}}^{\star 0} & \omega _{S}%
\end{array}%
\right) ^{\mu }{\normalsize ,}\;\scalefont{0.81}A^{\mu }=\frac{1}{\sqrt{2}}%
\left( 
\begin{array}{ccc}
\frac{f_{1N}+a_{1}^{0}}{\sqrt{2}} & a_{1}^{+} & K_{1}^{+} \\ 
a_{1}^{-} & \frac{f_{1N}-a_{1}^{0}}{\sqrt{2}} & K_{1}^{0} \\ 
K_{1}^{-} & {\bar{K}}_{1}^{0} & f_{1S}%
\end{array}%
\right) ^{\scalefont{0.81}\mu } \normalsize \label{LR}
\end{equation}%
and $\Delta =\mathrm{diag}(\delta _{N},\delta _{N},\delta _{S})$ describes
the explicit breaking of the chiral symmetry in the (axial-)vector channel (in terms of masses of $u$, $d$ and $s$
quarks, $\delta _{N} \sim m_{u,d}^2$ and $\delta _{S} \sim m_{s}^2$; isospin symmetry for non-strange quarks has been assumed).
Explicit symmetry breaking in the (pseudo)scalar sector is described by Tr$%
[H(\Phi +\Phi ^{\dagger })]$ with the constant matrix $H=1/2\,\mathrm{diag}(h_{0N},h_{0N},$ $\sqrt{2}h_{0S})$. Additionally, $D^{\mu}\Phi = \partial^{\mu}\Phi-ig_{1}(L^{\mu}\Phi-\Phi R^{\mu
})$ $-ieA^{\mu}[T_{3},\Phi]$ is the covariant
derivative; $L^{\mu \nu }=\partial ^{\mu }L^{\nu }-ieA^{\mu }[T_{3},L^{\nu
}]-\{ \partial ^{\nu }L^{\mu }$ $-ieA^{\nu }[T_{3},L^{\mu }] \} $ and $R^{\mu \nu }=\partial ^{\mu }R^{\nu }-ieA^{\mu }[T_{3},R^{\nu }]-\left\{ \partial ^{\nu
}R^{\mu }-ieA^{\nu }[T_{3},R^{\mu }]\right\} $ are, respectively, the
left-handed and right-handed field strength tensors, $A^{\mu }$ is the
electromagnetic field, $T_{3}$ is the third generator of the $SU(3)$ group
and the term $c_{1}(\det \Phi -\det \Phi ^{\dagger })^{2}$ describes the $%
U(1)_{A}$ anomaly \cite{Klempt}. \newline
We assign the field $\vec{\pi}$ to the pion; $\eta _{N}$ and $\eta_S$ are assigned, respectively, to the pure non-strange and the pure strange counterparts of the $\eta$ and $\eta^\prime$ mesons. The fields $\omega _{N}^{\mu }$, $\vec{\rho}^{\mu }$, $f_{1N}^{\mu }$
and $\vec{a}_{1}^{\mu }$ are assigned to the $\omega (782)$, $\rho (770)$, $%
f_{1}(1285)$ and $a_{1}(1260)$ mesons, respectively. We also
assign the $K$ fields to the kaons; the $\omega _{S}^{\mu }$, $%
f_{1S}^{\mu }$ and $K^{\star \mu }$ fields correspond to the $\varphi (1020)$%
, $f_{1}(1420)$ and $K^{\star }(892)$ mesons, respectively. Assignment of
the $K_{1}^{\mu }$ field is, unfortunately, not as clear since this state
can be assigned either to the $K_{1}(1270)$ or to the $K_{1}(1400)$
resonances. This is further discussed in Sec.\ \ref{Results}.\\
The isoscalar fields $\sigma _{N}$ and $\sigma _{S}$ mix in the Lagrangian (%
\ref{Lagrangian}) originating two mixed states; these states, together with
the non-strange isovector state $\vec{a}_{0}$ and the scalar kaon $%
K_{0}^{\star }$, can be assigned to resonances below or above 1 GeV in the
physical spectrum \cite{PDG}; results from our model prefer the latter
assignment \cite{Doktorarbeit,Paper3,LesHouches,Stockholm}. As a consequence, we assign our $\vec{a}_{0}$ and $%
K_{0}^{\star }$ states to $a_{0}(1450)$ and $%
K_{0}^{\star }(1430)$, respectively.

Spontaneous chiral-symmetry breaking requires a shift of the fields $\sigma _{N}$ and $\sigma _{S}$ by their respective vacuum
expectation values $\phi _{N}$ and $\phi _{S}$. We then observe
that mixing terms containing axial-vectors and pseudoscalars and $K^{\star }$
and $K_{0}^{\star }$ arise in the Lagrangian; these are removed as described
in Refs.\ \cite{Doktorarbeit,Paper3,Madrid}. 
Subsequently, renormalisation coefficients need
to be introduced for the pseudoscalar fields and $K_{0}^{\star }$ (more details in Refs.\ \cite%
{Doktorarbeit,Paper3,Madrid}). 

Lagrangian (\ref{Lagrangian}) contains 14 parameters: $\lambda _{1}$, $%
\lambda _{2}$, $c_{1}$, $h_{0N}$, $h_{0S}$, $h_{1}$, $h_{2}$, $h_{3}$, $%
m_{0}^{2}$, $g_{1}$, $g_{2}$, $m_{1}$, $\delta _{N}$, $\delta _{S}$.
Parameters $h_{0N}$ and $h_{0S}$ are determined from the extremum condition
for the potential obtained from Eq.\ (\ref{Lagrangian}). Parameter $\delta
_{N}$ is set to zero throughout this paper because the explicit
breaking of the chiral symmetry is small in the non-strange quark sector. The other 11 parameters
are calculated from a global fit including 21 observables: $f_{\pi }$, $%
f_{K} $, $m_{\pi }$, $m_{K}$, $m_{\eta }$, $m_{\eta ^{\prime }}$, $m_{\rho }$%
, $m_{K^{\star }}$, $m_{\omega _{S}\equiv \varphi (1020)}$, $m_{f_{1S}\equiv
f_{1}(1420)}$, $m_{a_{1}}$, $m_{a_{0}\equiv a_{0}(1450)}$, $m_{K_{0}^{\star
}\equiv K_{0}^{\star }(1430)}$, $\Gamma _{\rho \rightarrow \pi \pi }$, $%
\Gamma _{K^{\star }\rightarrow K\pi }$, $\Gamma _{\phi \rightarrow KK}$, $%
\Gamma _{a_{1}\rightarrow \rho \pi }$, $\Gamma _{a_{1}\rightarrow \pi \gamma
}$, $\Gamma _{f_{1}(1420)\rightarrow K^{\star }K}$, $\Gamma _{a_{0}(1450)}$, 
$\Gamma _{K_{0}^{\star }(1430)\rightarrow K\pi }$ (data from Ref.\ \cite{PDG}%
). Note that the observables entering the fit allow us to determine only
linear combinations $m_{0}^{2}+\lambda _{1}(\phi _{N}^{2}+\phi _{S}^{2})$
and $m_{1}^{2}+h_{1}\left( \phi _{N}^{2}+\phi _{S}^{2}\right) /2$ rather
than parameters $m_{0}$, $m_{1}$, $\lambda _{1}$ and $h_{1}$ by themselves.
However, it is nonetheless possible to calculate axial-vector masses and decay
widths (explicit formulas in Refs.\ \cite{Doktorarbeit,Paper3}) -- see Sec.\ \ref{Results}.

\section{Fit Results} \label{Results}

\begin{table}[th]
\centering
\begin{tabular}{|c|c|c|}
\hline
Observable & Fit [MeV] & Experiment [MeV] \\ \hline
$f_{\pi}$ & $96.3 \pm0.7 $ & $92.2 \pm4.6$ \\ \hline
$f_{K}$ & $106.9 \pm0.6$ & $110.4 \pm5.5$ \\ \hline
$m_{\pi}$ & $141.0 \pm5.8$ & $137.3 \pm6.9$ \\ \hline
$m_{K}$ & $485.6 \pm3.0$ & $495.6 \pm24.8$ \\ \hline
$m_{\eta}$ & $509.4 \pm3.0$ & $547.9 \pm27.4$ \\ \hline
$m_{\eta^{\prime}}$ & $962.5 \pm5.6$ & $957.8 \pm47.9$ \\ \hline
$m_{\rho}$ & $783.1 \pm7.0$ & $775.5 \pm38.8$ \\ \hline
$m_{K^{\star}}$ & $885.1 \pm6.3$ & $893.8 \pm44.7$ \\ \hline
$m_{\phi}$ & $975.1 \pm6.4$ & $1019.5 \pm51.0$ \\ \hline
$m_{a_{1}}$ & $1186 \pm6$ & $1230 \pm62$ \\ \hline
$m_{f_{1}(1420)}$ & $1372.5 \pm5.3$ & $1426.4 \pm71.3$ \\ \hline
$m_{a_{0}}$ & $1363 \pm1$ & $1474 \pm74$ \\ \hline
$m_{K_{0}^{\star}}$ & $1450 \pm1$ & $1425 \pm71$ \\ \hline
$\Gamma_{\rho\rightarrow\pi\pi}$ & $160.9 \pm4.4$ & $149.1 \pm7.4$ \\ \hline
$\Gamma_{K^{\star}\rightarrow K\pi}$ & $44.6 \pm1.9$ & $46.2 \pm2.3$ \\ 
\hline
$\Gamma_{\phi\rightarrow \bar{K}K}$ & $3.34 \pm0.14$ & $3.54 \pm0.18$ \\ 
\hline
$\Gamma_{a_{1}\rightarrow\rho\pi}$ & $549 \pm43$ & $425 \pm175$ \\ \hline
$\Gamma_{a_{1}\rightarrow\pi\gamma}$ & $0.66 \pm0.01$ & $0.64 \pm0.25$ \\ 
\hline
$\Gamma_{f_{1}(1420)\rightarrow K^{\star}K}$ & $44.6 \pm39.9$ & $43.9 \pm
2.2 $ \\ \hline
$\Gamma_{a_{0}}$ & $266 \pm12$ & $265 \pm13$ \\ \hline
$\Gamma_{K_{0}^{\star}\rightarrow K\pi}$ & $285 \pm12$ & $270 \pm 80$ \\ 
\hline
\end{tabular}%
\caption{Best-fit results for masses and decay widths compared with
experiment. Central values of observables are from Ref.\ \protect\cite{PDG}.
Errors of observables are considered according to the criterion max(5\%,
experimental error). The reason is that already isospin-breaking effects in
the physical hadron mass spectrum are of the order of 5\% [for instance the
difference between the charged and neutral pion masses or the masses of $a_1(1260)$ and $f_1(1285)$] but are completely neglected in our model.
We have therefore artificially increased the experimental errors to 5\% if the
actual error is smaller or used the experimental values if the error is
larger than 5\%. }
\label{Table1}
\end{table}

Results for the observables from our best fit are presented in Table \ref{Table1}.
The fit yields $\chi ^{2}=12.33$, i.e. $\chi ^{2}/(21$ observables $-11$
parameters$)=1.23$. Table \ref{Table1} demonstrates a remarkable
correspondence of our results with experimental data in all meson channels.
In particular, our fit yields $m_{a_{1}}=(1186\pm 6)$\ MeV, $\Gamma
_{a_{1}\rightarrow \rho \pi }=(549\pm 43)$\ MeV and $\Gamma
_{a_{1}\rightarrow \pi \gamma }=(0.66\pm 0.01)$ MeV. Note that $\Gamma _{a_{1}(1260)\rightarrow \pi
\gamma }=(0.64\pm 0.25)$ MeV \cite{PDG} is the
only experimental information regarding the $a_1(1260)$ meson which is not an estimate;
our result for $\Gamma _{a_{1}\rightarrow \pi
\gamma }$
is within the experimental interval and, being constrained by abundant
experimental input, it even produces an error for $\Gamma _{a_{1}(1260)\rightarrow \pi
\gamma }$ that is smaller than the one quoted by PDG. Thus results in the non-strange axial-vector channel
justify the interpretation of $a_1(1260)$ as (predominantly) a $\bar q q$ state.

Our fit allows us to calculate mass and decay width of the strange
axial-vector state $K_{1}$ as a prediction. We obtain $m_{K_{1}}=1282$
MeV, $\Gamma _{K_{1}\rightarrow K^{\star }\pi }=205$ MeV, $\Gamma
_{K_{1}\rightarrow \rho K}=44$ MeV and $\Gamma _{K_{1}\rightarrow \omega
K}=15$ MeV. The mass is very close to the PDG value $m_{K_{1}(1270)}=(1272\pm 7)$
MeV \cite{PDG}. However,
the full decay width is $264$~MeV, while the PDG data read $\Gamma
_{K_{1}(1270)}=(90\pm 20)$~MeV and $\Gamma _{K_{1}(1400)}=(174\pm 13)$~MeV.
Our result is therefore approximately three times too large when compared to
the data for $K_{1}(1270)$ and approximately $50\%$ too large when compared
to the data for $K_{1}(1400)$, even with errors omitted from the
calculation. These results demonstrate the necessity to include a
pseudovector $I(J^{PC})=1(1^{+-})$ nonet into our model and implement its
mixing with the already present axial-vector nonet.

\section{Conclusions} \label{Conclusions}

We have presented an extended Linear Sigma Model containing
(axial-)vector mesons (eLSM). We have performed a global fit of masses and decay
widths from which we have drawn two conclusions: (\textit{i}) the
non-strange axial-vector meson\ $a_{1}(1260)$ can be accommodated as a $\bar{%
q}q$ state within our model and (\textit{ii}) a correct description of the
strange axial-vector states $K_{1}(1270)$ and $K_{1}(1400)$ requires the
implementation of mixing between an axial-vector and a pseudovector nonet.
The latter also represents an outlook for further investigation of axial-vector mesons;
the model can, however, also be applied to further study other mesons in vacuum and
at finite temperatures and densities (see Refs.\ \cite{Doktorarbeit,Paper3} for details).

\end{document}